\documentclass[copyright,creativecommons]{eptcs}
\providecommand{\event}{F-IDE 2015}

\usepackage[T1]{fontenc}
\usepackage[scaled=0.81]{beramono}
\usepackage{hyperref}

\usepackage[pdftex]{graphicx}
\usepackage{xcolor}
\usepackage{xspace}

\usepackage{listings}
\usepackage{eiffel}
\newcommand{\e}[1]{\mbox{\lstinline|#1|}}

\newcommand{\ap}{Au\-to\-Proof\xspace}
\newcommand{\eve}{\textsc{eve}\xspace}

\title{The \ap Verifier: Usability \\by Non-Experts and on Standard Code}

\author{Carlo A.\ Furia \qquad\qquad Christopher M.\ Poskitt \qquad\qquad Julian Tschannen
\institute{Chair of Software Engineering, Department of Computer Science, ETH Z\"{u}rich, Switzerland}
\email{firstname.lastname@inf.ethz.ch}
}

\def\titlerunning{The AutoProof Verifier: Usability by Non-Experts and on Standard Code}
\def\authorrunning{C.A. Furia, C.M. Poskitt, and J. Tschannen}

\begin{document}

\maketitle

\begin{abstract}
Formal verification tools are often developed by experts \emph{for} experts; as a result, their usability by programmers with little formal methods experience may be severely limited.
In this paper, we discuss this general phenomenon with reference to \ap: a tool that can verify the full functional correctness of object-oriented software. 
In particular, we present our experiences of using \ap in two contrasting contexts representative of non-expert usage. First, we discuss its usability by students in a graduate course on software verification, who were tasked with verifying implementations of various sorting algorithms. Second, we evaluate its usability in verifying code developed for programming assignments of an undergraduate course. The first scenario represents usability by serious non-experts; the second represents usability on ``standard code'', developed without full functional verification in mind. We report our experiences and lessons learnt, from which we derive some general suggestions for furthering the development of verification tools with respect to improving their usability.
\end{abstract}

\section{Introduction}

\ap is a flagship component of \eve \cite{TFNM11-SEFM11}, the Integrated Development
Environment that combines various verification tools we develop as
part of our research.  \ap is an auto-active~\cite{auto-active-LM} prover that works on
Eiffel programs annotated with full-fledged functional specifications
in the form of contracts (pre- and postconditions, class invariants,
and other kinds of annotations for verification such as frame
specifications).  While all the tools integrated in \eve were
developed with an eye towards usability and practicality, \ap mainly
targets expert users who can provide detailed annotations to reason
automatically about complex functional properties of object-oriented
programs. 

In previous work, we demonstrated that \ap is a powerful and flexible tool in that
it supported the verification of a variety of benchmarks (including algorithmic
problems and object-oriented design idioms)~\cite{tacas15}, as well as the proof of full-functional correctness of a realistic general-purpose data structure library~\cite{EB2}.  At the same time, we
remain interested in assessing \ap's usability in contexts that go
beyond those normally evaluated by research in verification, namely:
\begin{itemize}

\item Usability by users who are \emph{serious
    non-experts}~\cite{auto-active-LM};

\item Usability on \emph{standard} object-oriented code, \emph{not}
  written with verification in mind.

\end{itemize}
This paper describes our first experiences in these areas, outlines
the obstacles that stand in the way of broader use of \ap
and other similar tools integrated in verification environments, and suggests future work to address them.

\paragraph{Usability by non-experts.}

Students of our graduate course on ``Software Verification'' used \ap
to carry out part of their projects, consisting of the specification,
implementation, and verification of sorting algorithms.  The students
can be considered serious users: they devoted a significant amount of
time to learning and using \ap for the project, and received proper
training in deductive verification during the course.  At the same
time, they remain \emph{non-experts}: they had little previous experience
with auto-active verification, and were generally unaware of the
research challenges and state of the art in the area.
\autoref{sec:teach-verif-with} reports their experience, highlighting
the major issues they faced, and the lessons we learnt about making
\ap{} -- and auto-active tools in general -- friendlier for non-expert
users.

\paragraph{Usability on standard code.}

When a verification tool such as \ap is used by experts -- often the same persons who developed the tool in the first place --,  it is normally applied to code written and annotated with full functional verification in mind.
The experts know the idioms that are more amenable to verification -- requiring fewer annotations or leading to better performance -- thanks to their familiarity with the tool's inner workings.
In contrast, even specifying code that has been produced independently of verification may turn out to be a substantial challenge.
\autoref{sec:verify-stand-client} discusses our experience with applying \ap to the verification of a number of programming assignments of our ``Introduction to Programming'' undergraduate course.
In this case, we (the experts) apply verification to \emph{standard} code written prior to the verification attempt and independent of it, and go as far as possible without changing the implementation.
We report the major hurdles encountered, and the lessons learnt about how to attain more flexibility in applying auto-active verifiers to standard code.

\section{An overview of \ap} \label{sec:an-overview-ap}

\ap is a verifier of the functional correctness of programs written in the Eiffel programming language.
\ap follows the so-called \emph{auto-active} approach~\cite{auto-active-LM} to verification: users interact indirectly by providing annotations in the input source code; then, each invocation of \ap proceeds autonomously without user interaction until it provides feedback upon terminating.

The \emph{annotations} used by \ap serve different, complementary purposes.
Pre- (\e{require}) and postconditions (\e{ensure}) define the functional specification to be verified.
Class invariants supplement the specification by defining general consistency properties of objects, as well as by supporting methodology-specific annotations useful for reasoning about relations between objects.
Other kinds of annotations -- such as loop (in)variants and intermediate assertions -- are lower level in that they provide information that is, strictly speaking, redundant but essential to guide verification to success.
For example, a loop invariant characterizes the semantics of a loop in a way that is amenable to reasoning about the effects of the loop within its context.

Following a standard approach in deductive verification, \ap uses the information in the source code and its annotations to encode \emph{verification conditions}: logic formulae whose validity entails the correctness of the input against its specification.
Like other auto-active verifiers (such as Spec\#~\cite{SpecSharp}, Dafny~\cite{Dafny}, and VCC~\cite{VCC}), \ap does not generate verification conditions directly, but encodes the semantics of Eiffel and its annotations into a Boogie program~\cite{BoogieManual}, which the Boogie verifier can process to generate the actual verification conditions and to submit them to the theorem prover Z3 to check for validity.
\ap picks up Boogie's output and translates it back to refer to the input Eiffel code.
Failed verification attempts, in particular, point to specific annotations that could not be verified by Boogie.

Providing informative feedback is a critical aspect in supporting usable auto-active verification.
\ap deals with very expressive -- hence, undecidable -- logics.
Therefore, when verification ``fails'' it may mean one of two things: either there is an error (that is, an inconsistency between the implementation and the specification); or the program is correct but the prover needs additional guidance (in the form of more detailed annotations) to complete a proof.
It may be hard for users to figure out which is the case, and where the error is or the additional annotations are needed.

\ap's verification techniques are geared towards object-oriented features; in particular, \ap supports \emph{semantic collaboration}~\cite{semicola}, a verification methodology that combines with \emph{ownership}~\cite{Leino04} to reason about complex object structures such as those that are idiomatic in object-oriented design.
Besides object-orientated features, \ap also fully supports algorithmic reasoning by means of \emph{model-based contracts}~\cite{PFM10-VSTTE10}.
This is a style of specification whereby annotations refer to mathematical \emph{models} such as sequences, sets, and bags; this supports abstraction in terms of a small number of entities that are purely applicative, and hence easy to express and reason upon in the underlying prover's logic.

\ap provides two main user interfaces.
It is fully integrated in \eve -- the Eiffel Verification Environment -- which provides a full-fledged IDE that integrates various development and verification tools.
A lightweight web-based interface is also available through ComCom\footnote{\url{http://cloudstudio.ethz.ch/comcom/\#AutoProof}}, where users can submit verification problems online.
The web interface is limited to verification of single classes, and does not offer much in the way of editing features such as auto-completion or even saving partial work; nonetheless, the possibility of running \ap in a web browser without having to install any software is quite convenient for non-advanced users.

Related work discusses \ap's techniques in more detail~\cite{tacas15}, and presents the results of applying it to verification challenges involving algorithms~\cite{sttt-paper} and realistic object-oriented software~\cite{EB2}.

\section{Teaching verification with \ap} \label{sec:teach-verif-with}

We evaluated the usability of \ap for non-expert users through the take-home project of our graduate-level software verification course, which required students to verify the functional correctness of various sorting algorithms. After a brief overview of the course and project, we describe the students' results, their most common difficulties, and general lessons learnt from their experiences.

\subsection{Description of the course and project}

\paragraph{Course overview and demographics.}
Our ``Software Verification'' graduate course\footnote{Webpage of the latest iteration: \url{http://se.inf.ethz.ch/courses/2014b_fall/sv/}} at ETH Z\"{u}rich has run annually for several years.
The curriculum attempts to embrace and convey the diversity of verification approaches: topics span program logics, abstract interpretation, model checking, and testing. The course comprises a lecture and exercise programme that weaves fundamental subjects (e.g., Hoare logic, automata-based model checking) with more advanced research (e.g., separation logic for object-orientation, model checking real-time systems), and challenges the students to assess the trade-offs and connections between the different techniques.

In addition to developing a strong basis in theory -- assessed through a written exam (worth 70\% of the final grade) -- the course advocates \emph{practice}, predominantly through a take-home project (worth 30\%) that challenges students to fully verify various sorting algorithms using state-of-the-art verification tools. The project is intended to push students beyond isolated, simple exercises, and give them a chance to experience the process of verification all the way through to the end -- ``warts and all''.

Software Verification is offered within the software engineering theme of our Master's program, but it is also open to undergraduate and PhD students.
It is typically taken by 10--20 students with about 75\% of them Master's, 20\% Bachelor's, and 5\% PhD's.

\paragraph{Project description.}
In the 2013--14 iterations of the course, the project required students to implement a basic list data structure and sorting algorithms that operate upon it, and to specify and verify their functionality as completely as possible. We asked for this to be done twice: first, in \ap, which allows for verification to take place at the level of code; and second, in Boogie, which allows for more fine-grained control over the low-level details of the proof machinery. Students were permitted to work in teams of up to three people, and were required to submit, along with their annotated code, a report comparing the verification task in the two tools and highlighting the trade-offs from working at the two different levels of abstraction.

To get the students started, and to ensure comparable, fairly-graded projects, we provided them with a skeleton Eiffel class containing the signatures of routines that they needed to implement, specify, and verify. While the routine bodies for sorting were left completely blank, some of the simpler routines were given with partial specifications or implementations, so as to help the students become familiar with Eiffel syntax more quickly. Furthermore, we provided the necessary annotations relating to \ap's methodology for ownership and semantic collaboration, allowing the students to focus their verification efforts on the algorithms themselves and not the object structures (which were anyway here simple -- just one main class). 

The \e{sort} routine for them to implement required the combination of bucket sort with another sorting algorithm (merge sort in 2013; quick sort in 2014). The routine called the former algorithm on large bounded inputs (if the list contained a certain number of elements with a certain range of values), and the latter algorithm otherwise (small list length, large values, or both). Furthermore, bucket sort could recursively call the other sorting algorithm (merge sort or quick sort) on the sublists in its buckets. By combining various algorithms, the project served to highlight the modular approach to verification present in tools like \ap and Boogie. Verification of the combined algorithm thus boiled down to verifying the correctness of the two sorting algorithms: in particular, that they returned lists that were sorted permutations of their input.

\begin{lstlisting}
sort
	do
		if count >= Max_count // 2 and has_small_elements (array) then
			array := bucket_sort (array)
		else
			array := quick_sort (array)
		end
	ensure
		is_sorted (array)
		is_permutation (array.sequence, old array.sequence)
	end
\end{lstlisting}

Routines were specified with model-based contracts that abstracted the list to a mathematical sequence (\e{SEQUENCE}) of integers. The query \e{is_permutation} exemplifies the expressive style of reasoning that such contracts permit: two sequences (abstractions of lists) are evaluated to be permutations of each other if, after converting them to bags (multisets), they are object-equal (that is value-equal, denoted by $\sim$). (The annotations \e{functional} and \e{ghost} indicate to \ap that \e{is_permutation} is a declarative function used only in specifications.)

\begin{lstlisting}
is_permutation (a, b: SEQUENCE [INTEGER]): BOOLEAN
	note
		status: functional, ghost
	do
		Result := a.to_bag $\sim$ b.to_bag
	end
\end{lstlisting}

Since the project required only a single class, students were free to choose between using the lightweight web-based interface to \ap in ComCom, or using it through the \eve IDE. Similarly for Boogie, students could work within the \emph{rise4fun} web interface\footnote{\url{http://rise4fun.com/Boogie/}} or download the verifier itself from CodePlex.

\paragraph{Tool training.}
Students were given the project description four weeks into the course. Prior to this, the lectures introduced them to the fundamentals of Hoare-style reasoning, as well as to the basics of auto-active verification in \ap and Boogie. We accompanied the tool lectures with hands-on exercises, which started simplistically (verifying 2--3 lines of code), before gradually building up to more challenging examples adapted from international verification competitions. Beyond this material and the course assistant's support, students also had access to an \ap tutorial, manual, and code repository,\footnote{\url{http://se.inf.ethz.ch/research/autoproof/}} as well as to extensive Boogie documentation \cite{BoogieManual}.

\subsection{Verification results and difficulties}

\paragraph{Verification results.}
We provide some impressions and results from the \ap task of the 2014 course project. (We do not compare with the 2013 project, since the version of \ap used then lacked support for framing and model-based contracts, both of which full functional verification requires.)

Table \ref{tab:projectresults} displays the results of the project for the nine groups that submitted. Every group was able to implement, specify, and verify the basic API for list data structures (as we had hoped -- these were intended as a warm-up exercise). The core challenge -- verifying the two sorting algorithms -- led to more variety. Every group was able to implement the sorting routines, with only a few overcomplicating the task. Most groups were able to fully specify the routines, but three groups suffered from inconsistent specifications which led to vacuous verification (we discuss the problem more generally later in this section); two groups had isolated and fixable cases of it, but in the third group they were extensive and impossible to salvage. Six (resp.\ five) groups were successfully able to verify that quick sort (resp.\ bucket sort) returned a sorted permutation of the input. We include in these numbers three groups who relied on, but could not prove, the correctness of an ``obviously'' true lemma (of the kind that we exemplify in the following paragraph). Unsurprisingly, the greatest challenge appeared to be in verifying that the output was a permutation of the input; every group (other than the ones suffering from inconsistent specifications) could prove the weaker property that the output was \emph{some} sorted list.

\begin{table}
\setlength{\tabcolsep}{4pt}
	\centering
\begin{tabular}{lrrrr}
\multicolumn{1}{c}{\textsc{routine}}  &  \multicolumn{1}{c}{\textsc{implemented}} &  \multicolumn{1}{c}{\textsc{specified}}  &  \multicolumn{1}{c}{\textsc{full proof}} & \multicolumn{1}{c}{\e{is_sorted} \textsc{only}} \\
\hline
\e{add}, \e{extend}, \e{remove}, \dots & 9 & 9 & 9 & -- \\
\e{quick_sort}  & 9 & 8 & 6 & 2  \\
\e{bucket_sort} & 9 & 6 & 5 & 2 \\
\\
\end{tabular}
\caption{Project tasks that student groups were able to complete (out of a total of nine groups).}
\label{tab:projectresults}
\end{table}

\paragraph{Common difficulties.}
The most common difficulty came from \ap's expressive model-based contracts. On the one hand, students quickly grasped the abstractions model-based contracts provide, and were generally able to \emph{specify} full functional correctness (predicates \e{is_sorted} and \e{is_permutation}). On the other hand, many students struggled to \emph{verify} these specifications -- especially for permutations -- because the success of verification required, to some extent, knowledge of how the model-based contracts were encoded in Boogie, and such details were opaque to all but the most inquisitive of our non-expert users. Failing proofs (of correct code) were often fixable through the addition of lemmas or intermediate assertions; but these were occasionally non-obvious, non-trivial, and far removed from the original verification task and the level of abstraction of the source code. One group, for example, realised that \ap was getting stuck on a proof because it could not automatically deduce an obvious relationship between bags (multisets) and subsequences:\footnote{The expression \e{s.interval (x, y)} denotes \e{s}'s subsequence from index \e{x} to index \e{y}.}

\begin{lstlisting}
   lemma_extend_bag (b: BAG [INTEGER]; s: SEQUENCE [INTEGER]; i: INTEGER)
   	require
   		in_range: 1 <= i and i <= s.count
   		same_until_i: b = s.interval (1, i - 1).to_bag
   	do
   	ensure
   		extended_same: b.extended (s [i]) = s.interval (1, i).to_bag
   	end
\end{lstlisting}

\noindent For \ap to prove the lemma, it was enough to add an extra assertion concerning intervals only:

\begin{lstlisting}
   check s.interval (1, i - 1).extended (s [i]) = s.interval (1, i) end
\end{lstlisting}

\noindent which was obtained, essentially, through trial and error. This process of guesswork appeared to be prevalent across groups in their attempts to prove the permutation property, and was a main source of frustration. It resulted, for several submissions, in a bloated annotation overhead, as students added whatever checks and contracts related to the models that they could think of, in the hope that they would lead to \ap instantiating the missing Boogie axioms it needed to verify the code.

A less common (but arguably more serious) difficulty came from the fact that \ap will happily verify programs with inconsistent specifications; in particular, routines with preconditions that reduce to \e{False}. There are no pre-states satisfying such preconditions, meaning, vacuously, that every valid execution (there aren't any) establishes the postcondition. There are currently no facilities built in \ap to check if specifications or assertions contain a contradiction of this kind.

The students who introduced contradictions typically did so using the advanced annotations for ownership- and collaboration-based reasoning. We had tried to prevent this by including all such annotations in the skeleton Eiffel class provided with the project description. Some groups, however, modified these contracts without truly understanding their semantics, leading to routines that vacuously verified. For example, one group added \e{array.is_wrapped} as a precondition to the \e{sort} routine; roughly, this means that \e{sort} operates on an \e{array} object that is not ``owned'' by any other object and is in a consistent state.
However, \e{sort}'s enclosing class includes clause \e{owns = [array]} as part of its invariant; this means that each instance of the class ``owns'' the \e{array} object it includes as attribute.
These two specification elements contradict each other: \e{array.is_wrapped} requires, in particular, that \e{array} is not owned by any object.

This general problem can be addressed on two levels: the tool can be improved to find inconsistent specifications automatically, and the students can be better educated to understand the shortcomings of \ap. Steps have been taken towards addressing the latter by adding a section about the problem to the \ap tutorial.
Supplying unit tests that exercise valid inputs can also help students understand the scope and flaws of their specifications; we plan to do this in future iterations of the course.
Adding support to \ap for finding inconsistencies belongs to future work, which could rely on and extend Boogie's ``smoke test'' feature.

\paragraph{User interface.}
As we discussed in \autoref{sec:an-overview-ap}, \ap offers two main user interfaces: \eve (a full-fledged IDE) and ComCom (a web-based lightweight interface).
Nearly all students preferred using the ComCom interface to carry out their projects.
The students generally appreciated the simplicity of ComCom's interface and the fact that they could use \ap through it on every computer without installing any software, and believed that such ease of use was overall more valuable than having additional tools provided by a full-fledged IDE.
This trade-off was clearly influenced by two features of the project work: the fact that the project was set up so as to involve developing all functionality in one single class (ComCom does not support multi-class projects); and the fact that we provided fairly complete online API documentation for the library classes needed for the project.
For more complex endeavors the support of an IDE could be invaluable; but for focused, algorithmic-verification projects such as the one we used, ease of use trumps having a rich toolset.

\section{Verifying standard client code with \ap} \label{sec:verify-stand-client}

The success of deductive verification -- in particular, of the auto-active fashion \ap conforms to (see~\autoref{sec:an-overview-ap}) -- often hinges on programs being written, specified, and annotated in a way that is amenable to automated reasoning.
Deductive verification hardly scales to poorly modularised programs, and certain language and design features may require more cumbersome annotations than other, semantically equivalent, features.
This is why showcase verification efforts (e.g.,~\cite{Why3-examples,VeriFast-examples,Dafny-examples,Bruns11,Mehnert12,gladischTyszberowicz2013,EB2}) normally target software that has been developed with verification in mind from the beginning, or at least has been significantly adapted to suit verification.

To evaluate how \ap behaves when these standard assumptions are not met, we tried to verify a number of programs that were not written with verification in mind.
This effort is indicative of \ap's usability on \emph{standard} code.
This section presents the programs we considered and how much we could verify with only limited changes to the source code before hitting the limitations of \ap.

\subsection{Description of the example programs}

We consider four programs that we used in our ``Introduction to Programming'' course.
The programs are three applications (simulating board games of increasingly richer logic and features) and one library (modelling a transportation system).
The applications are programming assignments for students of the course; we verified the master solutions developed by instructors in the past, which have the advantage of offering code of sufficiently good quality, but still not developed for verification.
The library is used -- but not written -- by the same students in other assignments; it was also developed by instructors of the course and improved over the years.

\begin{description}
\item[Board game 1 (\textsc{G1}):]
A number of players advance on a board made up of a fixed number of squares by rolling two dice; the first player that reaches the end of the board is the winner.
Players are identified by their names and positions on the board.
The board is represented only implicitly through the players' positions on it.

\item[Board game 2 (\textsc{G2}):]
Extending the logic of \textsc{G1}, players also collect money as they progress; the game ends as soon as a player reaches the end of the board; the player with the largest amount of money is the winner.
Players are identified by their names, positions on the board, and amount of money they hold.
The board is represented explicitly as a list of square objects, which are of three classes related by inheritance.
What happens when a player reaches a square depends on the class the square is an instance of: the player may win money, lose money, or neither win nor lose.

\item[Board game 3 (\textsc{G3}):]
A simplified version of Monopoly.
Players continue playing as long as they have money; the last player remaining with some money is the winner.
Square objects now include classes implementing more complex behaviour, such as properties that can be bought and sold by players.
Then, a player has to pay ``rent'' upon reaching a square representing property of another player.

\item[Traffic library (\textsc{TL}):]
The library supports the modelling of urban public transportation systems.
It offers classes modelling entities such as lines, stops, and carriers.
The classes use many cross-referencing data structures defining mutual consistency.
For example: stop classes contain a list of lines (lines that stop there); line classes contain a list of stops (where the line stops); and class invariants specify that, for each stop $s$, each line $\ell$ that is in $s$'s list of lines must include $s$ among its list of stops.
\end{description}

The design and implementation of the four programs significantly relies on common data structures (such as lists) taken from the EiffelBase2 library, the implementation of which we recently fully verified~\cite{EB2}.
With this setup, the effort described here focuses on the verification of standard client code that uses fully verified standard components.

\subsection{Verification process and results}

The four programs to be verified consist of fully compilable and executable implementations annotated with basic interface specifications in the form of preconditions, postconditions, and class invariants.
Our main goal was verifying the given specifications against the given implementations while changing them as little as possible.
We obviously had to add a significant number of annotations to support verification with \ap; we occasionally also had to extend or adjust existing specifications to make them amenable to automated reasoning.
We resorted to changing the implementation only when it was the only way to enable verification with \ap; in these cases, we limited ourselves to small local changes that ostensibly did not affect the programs' structure or behaviour.\footnote{Related work has looked into the problem of verifying systematic refactoring of code across versions~\cite{MitschQP14}.}

\paragraph{Changes necessary for verification.}
The most common changes we introduced were:

\begin{itemize}
\item
The given interface specifications were in general too weak for modular reasoning, where the effect of a call to routine $r$ is limited to what is prescribed by $r$'s pre- and postconditions (as well as the invariant of $r$'s enclosing class).
Thus, we added stronger specifications in the form of preconditions, postconditions, and class invariants.

\item
Frame specifications (denoted by the keyword \e{modify}) specify which locations in the heap each routine may modify.
Such specifications are essential for modular reasoning, but were not present in the original programs since Eiffel does not offer native syntax to specify them.
Thus, we added frame specifications to all routines to be verified.

\item
Relations between dependent objects (for example, an object of class \e{STOP} that includes a list object) require annotations following the ownership and semantic collaboration methodologies~\cite{semicola}.
Thus, we added annotations to this effect in the form of additional class invariants.

\item
Reasoning about loop correctness and termination requires loop invariants and variants, which were not given in the original programs.
Thus, we added loop invariants to prove postconditions and loop variants to prove termination.

\item
We replaced regular Eiffel strings with a custom \e{STRING} class annotated with model-based contracts.
This way, we were able to reason about basic string operations using \ap.

\item
The board games (precisely, the dice components) rely on random number generator functions implemented as \e{once} routines (Eiffel's variant of \verb+static+ methods in other languages).
\ap does not fully support \e{once} routines, as they require non-modular reasoning in general.
We removed \e{once} routines and mimicked their functionality using regular attributes.

\item
We added various intermediate assertions to guide \ap to successful verification without timeouts or spurious errors.

\end{itemize}

\autoref{tab:coderesults} displays data about the example programs.
Each program takes two rows: one row describes it \emph{before} we introduced the modifications necessary for verification, and another row describes it \emph{after} modification and verification. For each example, the table reports: the number of Eiffel classes (\textsc{\#C}) and routines (\textsc{\#R}); the lines of executable Eiffel \textsc{code} and of \textsc{annotations} (a total of $T$ annotation lines, split into preconditions $P$, postconditions $Q$, class invariants $C$, loop invariants and variants $L$, frame specifications $F$, intermediate assertions $A$, and auxiliary annotations $N$ specific to the verification methodology); the \textsc{a}/\textsc{c} annotations to code ratio (measured in tokens, as it is customary); and the percentage of verified routines $V$.

\begin{table}
\begin{center}
\setlength{\tabcolsep}{4pt}

\begin{tabular}{l rr r rrrrrrrr r@{.}l r}
\textsc{name} & \textsc{\#C} & \textsc{\#R} & \textsc{code} & \multicolumn{8}{c}{\textsc{annotations}} & \multicolumn{2}{c}{\textsc{a}/\textsc{c}} & \textsc{V}\\
&&&& {\scriptsize $T$} & {\scriptsize $P$} & {\scriptsize $Q$} & {\scriptsize $C$} &{\scriptsize $L$} & {\scriptsize $F$} & {\scriptsize $A$} & {\scriptsize $N$} & \multicolumn{2}{c}{} & \\

\hline

\textsc{G1} \hspace{2mm} \textit{before} & 
4 & 9 &
164 &
20 & 6 & 6 & 8 & 0 & 0 & 0 & 0 &
0&2
\\

\textsc{G1} \hspace{2mm} \textit{after} & 
4 & 8 &
165 &
101 & 17 & 13 & 16 & 37 & 3 & 3 & 12 &
1&2 &
100\%
\\
\hline

\textsc{G2} \hspace{2mm} \textit{before} & 
8 & 19 &
301 &
41 & 11 & 14 & 16 & 0 & 0 & 0 & 0 &
0&3 
\\

\textsc{G2} \hspace{2mm} \textit{after} & 
8 & 18 &
307 &
173 & 25 & 27 & 30 & 57 & 8 & 7 & 29 &
1&4 &
100\%
\\
\hline

\textsc{G3} \hspace{2mm} \textit{before} & 
15 & 38 &
491 &
87 & 19 & 31 & 37 & 0 & 0 & 0 & 0 &
0&2
\\

\textsc{G3} \hspace{2mm} \textit{after} & 
15 & 45 &
608 &
425 & 50 & 56 & 52 & 86 & 49 & 37 & 95 &
1&1 &
93\%
\\
\hline

\textsc{TL} \hspace{2mm} \textit{before} & 
13 & 145 &
1219 &
442 & 176 & 166 & 100 & 0 & 0 & 0 & 0 &
0&5
\\

\textsc{TL} \hspace{2mm} \textit{after} & 
14 & 149 &
1220 &
1077 & 275 & 225 & 112 & 30 & 46 & 114 & 275 &
1&3 &
78\%
\\
\hline

\end{tabular}

\end{center}
\caption{Results of verifying the programming examples.}
\label{tab:coderesults}
\end{table}

\paragraph{Verification results.}
Unsurprisingly, the success of our verification effort varied with the complexity of the example programs.

\textbf{Board games 1 and 2.}
We verified the correctness of each data structure usage according to their complete specifications in EiffelBase2; we also verified the functional correctness against the given specifications (augmented as discussed above).\footnote{The verified code of \textsc{G1} and \textsc{G2} is available online at \url{http://se.inf.ethz.ch/research/autoproof/repo}}
Overall, the verification of these two example programs was successful with reasonable effort and only minimal changes to the implementations.

The hardest and most time-consuming task was providing suitable loop invariants.
Note that the original versions of the programs did not include any loop invariants (see column $L$ in \autoref{tab:coderesults}).
The more complex the postconditions to be proved, the trickier the loop invariants were to determine.
In a few cases we had to introduce small changes to the initialisation, or to the order of instructions in the loops, so that we could write less complex loop invariants that \ap could reason about without becoming bogged down.
For example, we explicitly initialised variables before entering a loop -- even if the loop body assigns to the variable before reading it -- so that the loop invariant does not have to handle ``initiation'' (that is, the fact that the invariant holds initially before entering the loop for the first iteration) as a special case.

Reasoning using class invariants also required a good deal of additional annotations.
By default, \ap makes the whole class invariant of valid objects available wherever they are visible.
The example programs involve numerous data structures and other classes with complex invariants; hence, the default approach means that large, complicated assertions often cluttered the proof space.
Instead, we used \ap with a different option: no class invariants were visible by default and, whenever a specific clause of a specific class invariant was needed in the proof, we explicitly made it available to the prover with an \e{assert} annotation.
This option gave us much more flexibility, and supported the verification of routines involving objects with complex invariants, but also required significantly more annotations on a case-by-case basis.

\textbf{Board game 3.}
We verified the correctness of each data structure usage according to their complete specifications in EiffelBase2; we also verified functional correctness against the given specifications (augmented as discussed above) for a large part (84\%) of the routines.
Compared to board games 1 and 2, board game 3 required more substantial changes to the implementation and, generally, more effort.

Expressing and reasoning about class invariants -- in particular, framing specifications in combination with inheritance and the consistency of inter-dependent objects -- was the main source of complexity, and what prevented us from verifying all the routines.
Specifically, we failed to verify routines related to the class \e{PROPERTY} that models board squares in the style of Monopoly.
When a player lands on one such square, the player owning the property receives an amount of money as rent.
This behaviour cannot be expressed in the framing specification of \e{PROPERTY}'s parent class, which is more abstract and has no notion of players' property; introducing the behaviour directly in \e{PROPERTY} is hard to achieve without contradicting the parent's more abstract specification.
In addition, the specifications of classes \e{PROPERTY} and \e{PLAYER} are inter-dependent (each player has a list of properties, and each property has a player who owns it), which makes verification of these routines even more intricate.

One specific design issue which required some changes to the implementation was the usage of creation procedures (constructors).
Many classes relied on the default creation procedure, which simply initialises attributes to their default values (for example, \e{Void} for references).
\ap's methodology, however, introduces some special attributes to encode the relationship between dependent objects (ownership and collaboration relations); hence, we had to provide specialised creation procedures that initialise such special attributes in a way that enables verification.

\textbf{Traffic library.}
While we were still able to verify a substantial part of Traffic's routines, the effort required -- both in terms of developing the annotations and in the actual verification time -- was significantly higher.
Traffic's design consists of a core class containing several data structures holding all objects in the transportation system, whose class invariants express consistency between them.
This centralised design makes verification less modular, which is reflected by the longer verification times of routines that deal with multiple data structures (such as initialisation routines), even though their size in number of instructions is not substantially different from that of some complex routines from the board game examples.

As with \textsc{G3}, specifying classes and routines that rely on the consistency of other objects was a delicate and time-consuming task.
Complex operations -- for example, removing a station while ensuring that all cross references from the corresponding lines are removed, or generating a list of all lines connecting two stations ordered by name -- required major effort in providing specifications suitable for \ap (in particular, careful manipulation of ghost code to support efficient reasoning about model-based specifications), to the point that we could not complete a number of them within the guidelines. 

One specific issue that required changes to the implementation of the Traffic library was its usage of  floating point arithmetic for 2D vector operations and calculating distances between stations.
\ap has limited support for floating point operations, and translates floating point numbers to Boogie's \verb+real+ type (corresponding to mathematical reals).
Hence, we weakened postconditions of operations involving floating point arithmetic to accommodate the way that they are modelled by the prover.

\section{Related work} \label{sec:related-work}

Over the last decade or so, following the steady progress of verification tools, there has been an increasing interest in teaching formal methods using practical tools.
The TFM~\cite{tfm-2004,tfm-2009} and FORMED~\cite{formed} workshops present a variety of experiences based on approaches as diverse as model checking, test-case generation, and interactive proofs.
For brevity, this section focuses on few recent works that are closer to our own experience.

Poll~\cite{Poll09a} discusses teaching formal methods using Java, JML annotations, and the static verifier ESC/Java2.
During exercise sessions lasting 2--3 hours, students solve verification problems that are given by the instructors and involve different aspects of specification and verification.
Some of the findings are common to our experience: for example, how the students are initially baffled by how modular verification works (and are frustrated by having to provide detailed specifications of every used routine); or the problems resulting when verification trivially succeeds simply because the students inadvertently introduce inconsistent specifications.
Other findings reported by Poll are less relevant to our experience mainly because of the different setup.
\ap is a more modern tool than ESC/Java2 and it is sound; hence, it supports verification of fairly complex algorithmic challenges.
The tricky semantics of class invariants was not a major problem for us, since we deemphasised invariant reasoning in the project descriptions; in addition, the fact that students developed both implementations and specifications assuaged the challenge of dealing with specification styles that they were not familiar with.
However, it was a challenge when we tried to apply \ap to code developed without invariant reasoning (or, generally, formal verification) in mind.
Finally, the limited and incomplete API specifications of Java libraries were a problem for Poll's students; in our case, we provided the necessary library classes with detailed specifications so that students could focus on the specifications and code in their project modules.

Kiniry and Zimmerman~\cite{Kiniry-Zimmerman08a} advocate an early and gradual introduction of formal methods into the undergraduate curriculum, in conjunction with design by contract and runtime as well as static checking.
They emphasise the importance of tool support and IDE integration; our experience led to similar recommendations in this respect.
Their approach to object-oriented design and implementation has a broader scope than our project's; but we target more advanced functional properties that better match the capabilities of our \ap tool.

Jaume and Laurent~\cite{Jaume-Laurent14a} discuss the role of tool support -- and IDE integration -- to teach discrete mathematics concepts such as those that are ubiquitous in formal method specifications.
Their approach to relating mathematical structures is similar to the way mathematical structures are related by inheritance in our model classes.
They support proofs of theorems using the FoCaLiZe system, which relies on proof scripts provided by users.
This is suitable for teaching discrete mathematics and techniques such as induction; in contrast, \ap follows the auto-active and programs-as-proofs paradigm, which is closer to the source-code level of programs.

\section{Discussion} \label{sec:disc-future-work}

The fact that \ap lacks techniques to detect inconsistent specifications betrays its origins as a research tool ``by experts for experts'' -- given to serious non-expert users only a posteriori.
As expert users, we normally deploy several means of detecting inconsistency in specifications: manually inserting \e{assert False} instructions in suspicious locations, using Boogie's smoke tests on the Boogie translations \ap produces in the background, or even running Z3's axiom profiler to follow the instantiations of axioms during proofs.
We extended \ap's tutorial to suggest the first technique, which can be carried out at the level of source code; but part of the problem remains due to the students' inexperience: while we have learnt to be suspicious of complex proofs that suddenly succeed in no time (they're probably due to inconsistencies), students can get a false sense of progress from such behaviour and do not necessarily suspect that their effortless success is too good to be true.

We have been moderately successful at verifying standard code not written with verification in mind.
Our effort has two main sides, which highlighted clearly different challenges.
On one hand, verifying \emph{client usages} -- every call to library routines conforms to the library's API specification -- generally requires reasonable effort.
In our examples, the library classes were mainly data structures from EiffelBase2, with preconditions typically requiring that references be non-\e{Void} and position indexes be within bounds.
With few exceptions, such properties can be verified with moderate annotation overhead that scales graciously with the client program's size.
It is interesting to note that such properties are also amenable to fully automated techniques, such as static analysis, that require little or no annotation.

On the other hand, verifying \emph{functional correctness} of the actual programs was significantly more challenging.
For the simpler examples, the biggest challenge was dealing with complex class invariants that clutter the proof space; we tackled the challenge by trading off some automation in exchange for the capability of selecting the few class invariants that were relevant to each part of the correctness proofs.
For the more complicated examples, which exhibited complex dependencies between objects, simply selecting the relevant class invariants was not enough.
In this case, \emph{framing} became the number one challenge, which we could tackle only partially.
Even if \ap supports semantic collaboration, which offers constructs to model complex object dependencies, specifying sophisticated framing relationships is simply not possible as an afterthought.
In all, invariant reasoning is quite sensitive to the complexity and size of the interacting components.
The state of the art is such that scalability is limited and possible only if software is designed with great attention to modularisation and with an idea of verification's specific requirements.

\paragraph{Acknowledgements.} This work has been supported in part by ERC Grant CME \#291389.
Sebastian Nanz contributed to the design of the verification project as one of the instructors of ``Software Verification''.

\bibliographystyle{eptcs}
\bibliography{aproof,fm_teaching,base2}

\end{document}